\documentclass[12pt]{article}
\usepackage{a4wide,epsf,rotate,float,cite}

\newcommand{\zet}{\zeta_3}
\newcommand{\zef}{\zeta_5}
\newcommand{\wh}{\widehat}
\newcommand{\nn}{\nonumber}

\newcommand{\as}{\alpha_{s}}

\newcommand{\IM}{\mbox{\rm Im}}
\newcommand{\eqn}[1]{(\ref{#1})}
\newcommand{\mev}{\mbox{\rm MeV}}
\newcommand{\gev}{\mbox{\rm GeV}}
\newcommand{\geu}{\gamma_{{\rm E}}}
\newcommand{\smvs}{\vbox{\vskip 8mm}}
\newcommand{\MSb}{{\overline{\rm MS}}}
\newcommand{\sfrac}[2]{\mbox{$\frac{#1}{#2}$}}

\newcommand{\newsection}[1]{\section{#1}\setcounter{equation}{0}}

\begin{document}

\begin{titlepage}
\begin{flushright}
{\small\sf IFIC/01-52} \\[-1mm]
{\small\sf FTUV/01-1015} \\[-1mm]
{\small\sf HD-THEP-01-37} \\[-1mm]
{\small\sf FZ-IKP(TH)-01-20} \\[15mm]
\end{flushright}

\begin{center}
{\LARGE\bf Light quark masses from scalar sum rules}\\[15mm]

{\normalsize\bf Matthias Jamin${}^{1,*}$, Jos\'e Antonio Oller${}^{2}$
 and Antonio Pich${}^{3}$} \\[4mm]

{\small\sl ${}^{1}$ Institut f\"ur Theoretische Physik, Universit\"at
           Heidelberg,} \\
{\small\sl Philosophenweg 16, D-69120 Heidelberg, Germany}\\
{\small\sl ${}^{2}$ Departamento de F\'{\i}sica, Universidad de Murcia,
           E-30071 Murcia, Spain}\\
{\small\sl ${}^{3}$ Departament de F\'{\i}sica Te\`orica, IFIC,
           Universitat de Val\`encia -- CSIC,}\\
{\small\sl Apt. Correus 22085, E-46071 Val\`encia, Spain} \\[20mm]
\end{center}

{\bf Abstract:}
In this work, the mass of the strange quark is calculated from QCD sum
rules for the divergence of the strangeness-changing vector current. The 
phenomenological scalar spectral function which enters the sum rule is
determined from our previous work on strangeness-changing scalar form factors
\cite{jop:01a}. For the running strange mass in the $\MSb$ scheme, we find
$m_s(2\,\gev)=99\pm 16\;\mev$. Making use of this result and the light-quark
mass ratios obtained from chiral perturbation theory, we are also able to
extract the masses of the lighter quarks $m_u$ and $m_d$. We then obtain
$m_u(2\,\gev)=2.9\pm 0.6\;\mev$ and $m_d(2\,\gev)=5.2\pm 0.9\;\mev$. In
addition, we present an updated value for the light quark condensate.

\vfill



\vspace{4mm}
{\small ${}^{*}$ Heisenberg fellow.}
\end{titlepage}


\newsection{Introduction}

Together with the strong coupling constant, quark masses are fundamental QCD
input parameters of the Standard Model, and thus their precise determination
is of paramount importance for present day particle phenomenology. In the light
quark sector, the mass of the strange quark $m_s$ deserves particular interest,
because its present uncertainty severely limits the precision of current
predictions of the CP-violating observable $\varepsilon'/\varepsilon$. The
ratios of light quark masses are known rather precisely from chiral
perturbation theory $\chi$PT \cite{gl:84,gl:85}, and thus, once the absolute
scale is set by $m_s$, also the masses of the lighter up and down quarks can
be determined.

Until today, two main methods have been employed to determine the strange quark
mass. QCD sum rules \cite{svz:79,rry:85,nar:89,shi:92} have been applied to
various channels containing strange quantum numbers, in particular the scalar
channel that will be the subject of this work \cite{nprt:83,jm:95,cdps:95,
cps:96,cfnp:97,jam:98,bgm:98,mal:99}, the pseudoscalar channel \cite{km:01},
the Cabibbo suppressed $\tau$-decay width \cite{pp:98,ckp:98,aleph:99,pp:99,
kkp:00,km:00,cdghpp:01}, as well as the total $e^+e^-$ cross section
\cite{nar:95,mal:98,mw:99,nar:99}. Also lower bounds on the strange mass have
been determined in the framework of QCD sum rules \cite{lrt:97,ynd:98,dn:98,
ls:98,km:01}. In addition, lattice QCD simulations for various hadronic
quantities have been used to extract the strange quark mass. For two recent
reviews where original references can be found, the reader is referred to
\cite{lub:00,gm:01}.

The dispersive QCD sum rule approach makes use of the phenomenological
knowledge on the spectral functions associated with hadronic currents with
the corresponding quantum numbers. From the experimental point of view at
present the cleanest information comes from $\tau$ decays \cite{aleph:99};
however, up to now the Cabibbo-suppressed hadronic $\tau$-decay data has not
been resolved into separate $J=0$ and $J=1$ contributions and the theoretical
uncertainties associated with a bad perturbative behaviour of its scalar
component put a limit on the achievable accuracy \cite{pp:98,aleph:99,pp:99,
kkp:00,km:00,cdghpp:01}.

The more standard analysis of the scalar or pseudoscalar currents provides a
large sensitivity to light quark masses. Unfortunately, the rather large
uncertainties of the $J=0$ data introduce important systematic errors in the
resulting quark mass determination, which are difficult to quantify. Previous
analyses have used phenomenological parameterisations based on saturation by
the lightest hadronic states with the given quantum numbers, sometimes improved
with Breit-Wigner and/or Omn\`es expressions \cite{nprt:83,jm:95,cdps:95,
cps:96,cfnp:97,jam:98,bgm:98,mal:99,km:01}.

In two recent papers, we have presented very detailed analyses of S-wave $K\pi$
scattering \cite{jop:00} and the $K\pi$, $K\eta$ and $K\eta'$ scalar form
factors \cite{jop:01a}, which incorporate the experimental knowledge on the
$J=0$ $K\pi$ phase shifts as well as all known theoretical constraints from
chiral perturbation theory, short-distance QCD, dispersive relations, unitarity
and large-$N_c$ considerations. The output of these works is a rather reliable
determination of the scalar spectral function up to about $2\;\gev$. This
allows us to perform a considerable step forward in the QCD sum rule
determination of light quark masses through the scalar correlators.

The central object which is investigated in the original version of QCD sum
rules \cite{svz:79} is the two-point function $\Psi(p^2)$ of two hadronic
currents
\begin{equation}
\label{psiq2}
\Psi(p^2) \;\equiv\; i \int \! dx \, e^{ipx} \,
\langle\Omega| \, T\{\,j(x)\,j(0)^\dagger\}|\Omega\rangle\,,
\end{equation}
where $\Omega$ denotes the physical vacuum and in our case $j(x)$ will be the
divergence of the strangeness-changing vector current,
\begin{equation}
\label{eq:1.2}
j(x) \;=\; \partial^\mu (\bar s\gamma_\mu q)(x) \; = \;
i\,(m_s-\hat m)(\bar s\,q)(x) \,.
\end{equation}
Since we work in the isospin limit, $q$ can be either an up- or down-type
quark, and $\hat m$ is the isospin average $\hat m=(m_u+m_d)/2$. To a good
approximation, $\Psi(p^2)$ is thus given by $m_s^2$ times the two-point
function of the scalar current.

Up to a subtraction polynomial, $\Psi(p^2)$ satisfies a dispersion relation,
\begin{equation}
\label{disrel}
\Psi(p^2) \; = \; \Psi(0) + p^2\,\Psi'(0) + p^4\!\int\limits_0^\infty
\frac{\rho(s)}{s^2(s-p^2-i0)}\,ds \,,
\end{equation}
where $\rho(s)\equiv\IM\,\Psi(s+i0)/\pi$ is the spectral function corresponding
to $\Psi(s)$. To suppress contributions in the dispersion integral coming from
high invariant-mass states, it is convenient to apply a Borel (inverse Laplace)
transformation to eq.~\eqn{disrel} \cite{svz:79}, which furthermore removes the
subtractions. The left-hand side of the resulting equation is calculable in QCD,
whereas under the assumption of quark-hadron duality, the right-hand side can
be evaluated in a hadron-based picture, thereby relating hadronic quantities
to the fundamental QCD parameters.

Generally, however, from experiments the phenomenological spectral function
$\rho_{ph}(s)$ is only known from threshold up to some energy $s_0$. Above this
value, we shall use the perturbative expression $\rho_{th}(s)$ also for the
right-hand side. This is legitimate if $s_0$ is large enough so that
perturbation theory is applicable. The central equation of our sum-rule
analysis for $m_s$ is then:
\begin{equation}
\label{sr}
u\,{\cal B}_u\Big[\Psi_{th}(p^2)\Big] \;\equiv\; u\,\wh\Psi_{th}(u) \;=\;
\int\limits_0^{s_0} \rho_{ph}(s)\,e^{-s/u}\,ds + \int\limits_{s_0}^\infty
\rho_{th}(s)\,e^{-s/u}\,ds \,,
\end{equation}
where ${\cal B}_u$ is the Borel operator, the hat denotes the Borel
transformation, and $u$ is the so-called Borel variable. The main ingredients
in this equation, namely the theoretical expression for the two-point function
as well as the phenomenological spectral function, will be discussed
below.\footnote{Further details on the approach can for example be found in
ref.~\cite{jm:95}.}

In addition, it is instructive to investigate the sum rule which arises by
considering the Borel transform of $\Psi(p^2)/p^2$:
\begin{equation}
\label{sr0}
u\,{\cal B}_u\Biggl[\frac{1}{p^2}\Psi_{th}(p^2)\Biggr] \;\equiv\;
\wh\Phi_{th}(u) \;=\; \int\limits_0^{s_0} \frac{ds}{s}\,\rho_{ph}(s)\,
e^{-s/u} + \int\limits_{s_0}^\infty \frac{ds}{s}\,\rho_{th}(s)\,e^{-s/u} -
\Psi(0) \,.
\end{equation}
The sum rule \eqn{sr0} is constructed such that the subtraction constant
$\Psi(0)$ remains. However, from a Ward identity \cite{bro:81} this constant
is related to the following product of quark masses and quark condensates:
\begin{equation}
\label{ward}
\Psi(0) \;=\; (m_s-\hat m)\Big(\langle\Omega|\bar qq|\Omega\rangle -
\langle\Omega|\bar ss|\Omega\rangle\Big) \,.
\end{equation}
Note that the quark condensates in eq.~\eqn{ward} appear as non-normal-ordered
vacuum averages, and thus $\Psi(0)$ is not renormalisation group invariant
\cite{sc:88,jm:93,cdps:95,jam:02}. The corresponding renormalisation invariant
quantity involves additional quartic quark mass terms \cite{sc:88}. Because
of the dependence on $\Psi(0)$, analysing the sum rule of eq.~\eqn{sr0} would
enable us to obtain information on the quark condensates. As we shall show in
the next section, however, the perturbative expansion for $\wh\Phi_{th}(u)$
behaves very badly, and thus such an analysis appears to be questionable.
Additional discussion of $\Psi(0)$ can also be found in ref.~\cite{jam:02}.

In the next two sections, we present expressions for the theoretical as well as
phenomenological two-point functions which are relevant for the sum rules under
investigation. In section 4, we then discuss the numerical analysis of the
strange mass sum rule. Finally, in our conclusions, we compare our results with
other recent determinations of $m_s$, calculate the light-quark masses $m_u$
and $m_d$ from mass ratios known from $\chi$PT, and update our current
knowledge of the quark condensate.

\newsection{Theoretical two-point function}

In the framework of the operator product expansion the Borel transformed
two-point function $\wh\Psi_{th}(u)$ can be expanded in inverse powers of
the Borel variable $u$:
\begin{equation}
\label{psihat}
\wh\Psi_{th}(u) \;=\; (m_s-\hat m)^2\,u\,\biggl\{\,\wh\Psi_0(u)+\frac{\wh\Psi_2
(u)}{u}+\frac{\wh\Psi_4(u)}{u^2}+\frac{\wh\Psi_6(u)}{u^3}+\ldots\,\biggr\} \,.
\end{equation}
The $\wh\Psi_n(u)$ contain operators of dimension $n$, and their remaining $u$
dependence is only logarithmic. Below, we shall review explicit expressions
for the first two of these contributions.

The purely perturbative contribution $\wh\Psi_0(u)$ is presently known up to
${\cal O}(\as^3)$ \cite{bnry:81,gkls:90,che:96} and the expansion in the strong
coupling up to this order reads
\begin{eqnarray}
\label{psihat0}
\wh\Psi_0(u) &\!\!=\!\!& \frac{3}{8\pi^2}\,\biggl[\,1+\Big(\sfrac{11}{3}+2\geu
\Big)a+\Big(\sfrac{5071}{144}-\sfrac{17}{24}(\pi^2-6\geu^2)+\sfrac{139}{6}
\geu-\sfrac{35}{2}\zet\Big)a^2+\Big(\sfrac{1995097}{5184}-\sfrac{\pi^4}{36}
\nn \\
\smvs
& & -\,\sfrac{695}{48}(\pi^2-6\geu^2)-\sfrac{221}{48}\geu(\pi^2-2\geu^2)+
\sfrac{2720}{9}\geu-\sfrac{475}{4}\geu\zet-\sfrac{61891}{216}\zet+
\sfrac{715}{12}\zef\Big)a^3\,\biggr] \nn \\
\smvs
&\!\!=\!\!& \frac{3}{8\pi^2}\,\Big[\,1+1.535\,\as+2.227\,\as^2+1.714\,\as^3\,
\Big] \,,
\end{eqnarray}
where $a\equiv\as/\pi$, $\geu$ is Euler's constant and $\zeta_z\equiv\zeta(z)$
is the Riemann $\zeta$-function. In this expression the logarithmic corrections
have been resummed to all orders, and thus the strong coupling $\as(u)$ should
be evaluated at the scale $u$. Higher order terms are also known in the
large-$N_f$ expansion \cite{bkm:00} and partial results are known at order
$\as^4$ \cite{bck:01}. Even for $\as(1\,\gev)\approx 0.5$ the last term in
\eqn{psihat0} is only about 20\% and the perturbative expansion displays a
reasonable convergence. Because the two-point function scales as $m_s^2$, the
resulting uncertainty for $m_s$ from higher orders is at most 10\%. In practice
it is much smaller since the average scale at which the sum rule is evaluated
lies around $1.5\,\gev$.

The theoretical two-point function $\wh\Phi_{th}(u)$ of eq.~\eqn{sr0} has an
operator product expansion which is completely equivalent to eq.~\eqn{psihat},
and the corresponding perturbative contribution takes the form:
\begin{eqnarray}
\label{phihat0}
\wh\Phi_0(u) &\!\!=\!\!& \frac{3}{8\pi^2}\,\biggl[\,1+\Big(\sfrac{17}{3}+2\geu
\Big)a+\Big(\sfrac{9631}{144}-\sfrac{17}{24}(\pi^2-6\geu^2)+\sfrac{95}{3}
\geu-\sfrac{35}{2}\zet\Big)a^2+\Big(\sfrac{4748953}{5184}-\sfrac{\pi^4}{36}
\nn \\
\smvs
& & -\,\sfrac{229}{12}(\pi^2-6\geu^2)-\sfrac{221}{48}\geu(\pi^2-2\geu^2)+
\sfrac{4781}{9}\geu-\sfrac{475}{4}\geu\zet-\sfrac{87541}{216}\zet+
\sfrac{715}{12}\zef\Big)a^3\,\biggr] \nn \\
\smvs
&\!\!=\!\!& \frac{3}{8\pi^2}\,\Big[\,1+2.171\,\as+5.932\,\as^2+17.337\,\as^3\,
\Big] \,.
\end{eqnarray}
As is obvious from this expression, the perturbative expansion for
$\wh\Phi_0(u)$ behaves very badly. Even at a scale $\sqrt{u}=2\;\gev$, the
last two terms are of comparable size and individually both are larger than
50\% of the leading term. If the logarithmic corrections are not resummed, the
perturbative expansion could be improved by taking a fixed scale $\mu$. This
would shift part of the corrections into the prefactor $(m_s-\hat m)^2$. In
this case one finds, however, that for $\sqrt{u}$ in the range $1-2\;\gev$ a
reasonable size of the higher orders is only obtained if $\mu$ is much less
than $1\;\gev$. But then the perturbative contribution is again questionable.
To conclude, the huge perturbative corrections for $\wh\Phi_0(u)$ prevent us
from performing a sum rule analysis of eq.~\eqn{sr0}.

The next term in the operator product expansion $\wh\Psi_2(u)$ only receives
contributions proportional to the quark masses squared. Its explicit expression
reads
\begin{equation}
\label{psihat2}
\wh\Psi_2(u) \;=\; -\,\frac{3}{4\pi^2}\,\biggl\{\,\Big[\, 1 + \sfrac{4}{3}
(4+3\geu)\,a\,\Big](m_s^2+m_u^2) + \Big[\, 1 + \sfrac{4}{3}(7+3\geu)\,a\,\Big]
m_s m_u \,\biggr\} \,.
\end{equation}
Already at a scale of $u=1\,\gev^2$ the size of $\wh\Psi_2$ is less than 3\%,
decreasing like $1/u$ for higher scales. Although it has been included in the
phenomenological analysis, for the error estimates on the strange quark mass
it can be safely neglected.

The same holds true for the dimension-four operators. In this case there are
contributions from the quark and gluon condensates as well as quark mass
corrections of order $m^4$. Again, at a scale of $u=1\,\gev^2$ the size of
$\wh\Psi_4$ is well below 1\% of the full two-point function, hence being
negligible for the strange mass analysis. Nevertheless, the dimension-four and
in addition also the dimension-six contributions $\wh\Psi_4$ and $\wh\Psi_6$
have been included in our numerical investigations. Analytic expressions for
these contributions are collected in section~2 of ref.~\cite{jm:95}.

To calculate the perturbative continuum on the right-hand side of eq.~\eqn{sr},
we also need the theoretical spectral function $\rho_{th}(s)$ which is given by
\begin{eqnarray}
\label{rhoth}
\rho_{th}(s) &\!\!=\!\!& \frac{3}{8\pi^2}\,(m_s-\hat m)^2 s\,\biggl[\, 1 +
\sfrac{17}{3}\,a + \Big(\sfrac{9631}{144}-\sfrac{35}{2}\zet-\sfrac{17}{12}\pi^2
\Big)a^2 \nn \\
\smvs
& & \hspace{36.6mm} +\,\Big(\sfrac{4748953}{5184}-\sfrac{91519}{216}\zet+
\sfrac{715}{12}\zef-\sfrac{229}{6}\pi^2-\sfrac{\pi^4}{36}\Big)a^3 \,\biggr] \\
\smvs
&\!\!=\!\!& \frac{3}{8\pi^2}\,(m_s-\hat m)^2 s\,\Big[\,1+1.804\,\as+3.228\,
\as^2+2.875\,\as^3\,\Big] \,. \nn
\end{eqnarray}
Again, the logarithms have been resummed, so that the coupling and masses are
running quantities evaluated at the scale $s$. It is possible to calculate the
relevant integral from $s_0$ to infinity in eq.~\eqn{sr} analytically. The
corresponding theoretical expressions can be found in ref.~\cite{jm:95}.

\newsection{Hadronic spectral function}

The phenomenological spectral function is obtained by inserting a complete set
of intermediate states $\Gamma$ with the correct quantum numbers in the current
product of eq.~\eqn{psiq2},
\begin{equation}
\label{rhoph}
\rho_{ph}(s) \;=\; (2\pi)^3 \sum\limits_\Gamma\hspace{-5mm}\int\;
|\langle\Omega|j(0)|\Gamma\rangle|^2 \delta(p-p_\Gamma) \,,
\end{equation}
where $s=p^2$ and the integration ranges over the phase space of the hadronic
system with momentum $p_\Gamma$. In the case of the strangeness-changing scalar
current, the lowest lying state which contributes in the sum is the
$K\pi$-system in an S-wave isospin-$1/2$ state.

Including also the $K\eta$ and $K\eta'$ states, the scalar spectral function
can be written as
\begin{equation}
\label{rhoS}
\rho_{ph}(s) \;=\; \frac{3\Delta_{K\pi}^2}{32\pi^2}\,\biggl[\,
\sigma_{K\pi}(s)|F_{K\pi}(s)|^2 + \sigma_{K\eta}(s)|F_{K\eta}(s)|^2 +
\sigma_{K\eta'}(s)|F_{K\eta'}(s)|^2 \,\biggr] \,,
\end{equation}
with $\Delta_{K\pi}\equiv M_K^2-M_\pi^2$. The two-particle phase space factors
 $\sigma_{KP}(s)$ take the form
\begin{equation}
\label{sigma}
\sigma_{KP}(s) \;=\; \theta\Big(s-(M_K+M_P)^2\Big)\,\sqrt{ \Big(1-
\sfrac{(M_K+M_P)^2}{s}\Big) \Big(1-\sfrac{(M_K-M_P)^2}{s}\Big)} \,,
\end{equation}
where $P$ corresponds to one of the states $\pi$, $\eta$ or $\eta'$, and
the strangeness-changing scalar form factors $F_{KP}(s)$ are defined by
\begin{equation}
\label{FS}
\langle\Omega |\partial^\mu (\bar s \gamma_\mu u)(0) | KP \rangle \;\equiv\;
-\,i\,\sqrt{\frac{3}{2}}\;\Delta_{K\pi}\,F_{KP}(s) \,.
\end{equation}

Experimentally, it has been shown that the S-wave isospin-$1/2$ $K\pi$ system
is elastic below roughly $1.3\;\gev$, and below $2\;\gev$, $K\eta'$ is the
dominant inelastic channel \cite{est:78,ast:88}. Thus including these two states
should give a good description of the scalar spectral function below $2\;\gev$.
For completeness, however, in eq.~\eqn{rhoS} we have also taken into account
the $K\eta$ state. Multiparticle states, the lightest of which is the
$|K\pi\pi\pi\rangle$ state, have been neglected in \eqn{rhoS}. Theoretically,
their contributions are suppressed both in the chiral and large-$N_c$
expansions. Nevertheless, since at an energy around $2\;\gev$ they should play
some role, we intend to investigate these contributions in the future. Owing to
the positivity of the scalar spectral function, these additional contributions
should slightly increase the value of the strange quark mass.

In our previous work \cite{jop:01a}, the scalar form factors $F_{KP}(s)$
have been determined for the first time from a dispersive coupled-channel
analysis of the $K\pi$ system. As an input in the dispersion integrals, S-wave
$KP$ scattering amplitudes were used which had been extracted from fits to
the $K\pi$ scattering data \cite{est:78,ast:88} in the framework of unitarised
$\chi$PT with explicit inclusion of resonance fields \cite{jop:00}. The fact
that the $K\eta$ channel only gives a negligible contribution to the hadronic
spectral function was also corroborated in \cite{jop:01a}. Therefore, making
use of the results of ref.~\cite{jop:01a}, we are in a position to provide the
scalar spectral function in an energy range from threshold up to about
$2\;\gev$.

\newsection{Numerical analysis}

\begin{figure}[htb]
\centerline{
\rotate[r]{
\epsfysize=15cm\epsffile{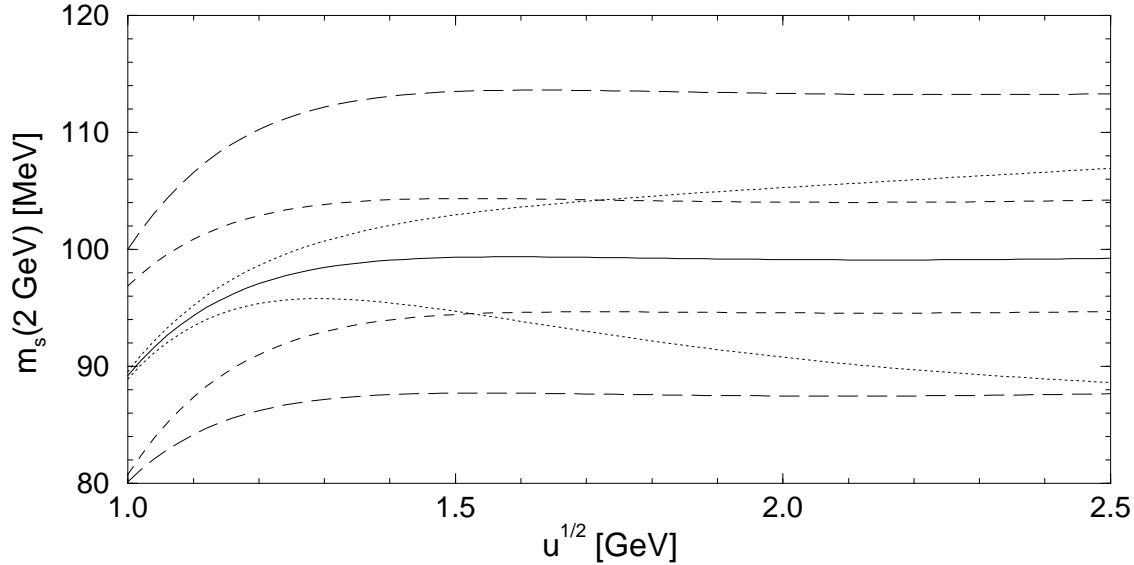} } }
\vspace{-4mm}
\caption[]{The strange mass $m_s(2\,\gev)$ as a function of $\sqrt{u}$.
Solid line: central parameters; long-dashed lines: (6.10K4) with $F_{K\pi}
(\Delta_{K\pi})=1.23$ (upper line), (6.10K3) with $F_{K\pi}(\Delta_{K\pi})=
1.21$ (lower line); dashed lines: $\alpha_s(M_Z)=0.1205$ (lower line),
$\alpha_s(M_Z)=0.1165$ (upper line); dotted lines: $s_0=4.2\;\gev^2$ (upper
line), $s_0=5.8\;\gev^2$ (lower line).
\label{fig1}}
\end{figure}

Evaluating the sum rule of eq.~\eqn{sr} with the theoretical two-point function
of section~2 and the hadronic spectral function of section~3, the resulting
values for the running strange quark mass $m_s(2\,\gev)$ as a function of
$\sqrt{u}$ are displayed in figure~\ref{fig1}. The solid line corresponds to
central values for all input parameters and constitutes our main result. For
$\hat m$, we have used $\hat m(2\,\gev)=4.05\;\mev$ which arises from our
analysis of the next section. From the region of maximal stability of the sum
rule (the extremum) which lies in the region of the $K_0^*(1430)$ resonance,
we extract our central value for the strange mass $m_s(2\,\gev)=99.4\;\mev$.
In the stability region, the continuum is only about 25\% of the full left-hand
side of eq.~\eqn{sr}, so that it should be under control. To give an estimation
of the uncertainties for $m_s$, let us discuss the inputs and their variation
in more detail.

The dominant source of uncertainty for $m_s$ is the hadronic spectral function.
To obtain an estimate of the corresponding error, we have calculated $m_s$
from different fits for the scalar form factors of ref.~\cite{jop:01a}. Since
the $K\eta$ channel was found to be unimportant, we have only considered the
two-channel spectral functions with contributions from $K\pi$ and $K\eta'$.
As the fits for the form factors, we utilise here our best fits (6.10K3) and
(6.10K4) of \cite{jop:01a}.\footnote{The strange mass resulting from the fit
(6.11K4) of \cite{jop:01a} is practically identical to $m_s$ from the fit
(6.10K4). Thus, for this work, we have not considered this fit separately.}
As was discussed in detail in ref.~\cite{jop:01a}, however, these fits are
not unique, but can be parametrised by $F_{K\pi}(\Delta_{K\pi})$ which should
take the value $1.22\pm0.01$. The solid line in figure~\ref{fig1} then
corresponds to the central value $F_{K\pi}(\Delta_{K\pi})=1.22$ and an average
of the spectral functions for (6.10K3) and (6.10K4). Varying $F_{K\pi}
(\Delta_{K\pi})$ for both fits, the largest $m_s$ is obtained for (6.10K4)
with $F_{K\pi}(\Delta_{K\pi})=1.23$ and the smallest for (6.10K3) with
$F_{K\pi}(\Delta_{K\pi})=1.21$. Both cases are displayed as the long-dashed
lines in figure~\ref{fig1} and the variation of $m_s$ has been collected in
table~\ref{tab1}.

\begin{table}[thb]
\renewcommand{\arraystretch}{1.2}
\begin{center}
\begin{tabular}{ccc}
\hline
Parameter & Value & $\Delta m_s$ [MeV] \\
\hline
$\rho_{ph}^{{\rm (6.10K4)}}(s)$ & $F_{K\pi}(\Delta_{K\pi})=1.23$ & $+14.3$ \\
$\rho_{ph}^{{\rm (6.10K3)}}(s)$ & $F_{K\pi}(\Delta_{K\pi})=1.21$ & $-11.6$ \\
$\alpha_s(M_Z)$ & $0.1185 \pm 0.0020$ & ${}^{+5.0}_{-4.7}$ \\
${\cal O}(\as^3)$ & ${}^{{\rm no}\;{\cal O}(\as^3)}_{2\times{\cal O}(\as^3)}$ &
${}^{+3.3}_{-3.6}$ \\
$s_0$ & $4.2 - 5.8\;\gev^2$ & ${}^{+4.3}_{-3.5}$ \\
\hline
\end{tabular}
\end{center}
\caption{Values of the main input parameters and corresponding uncertainties
for $m_s(2\,\gev)$. For a detailed explanation see the discussion in the text.
\label{tab1}} 
\end{table}

The next-largest uncertainty for $m_s$ which is related to the perturbative
expansion results from two sources. On the one hand there is an error on the
input value for $\as$ and on the other hand, there are unknown higher order
corrections. For the strong coupling, we have used the PDG value \cite{pdg:00}
and varied $\as$ within its error. The corresponding variation of $m_s$ is
shown as the dashed line in figure~\ref{fig1} where the upper line is the case
with $\alpha_s(M_Z)= 0.1165$ and the lower line with $\alpha_s(M_Z)=0.1205$.
To estimate the second uncertainty, we have either completely removed the
${\cal O}(\as^3)$ correction or doubled its value. The resulting errors for
$m_s$ are presented in table \ref{tab1}.

Another uncertainty for $m_s$ results from a variation of the continuum
threshold $s_0$. Our central value $s_0=4.75\;\gev^2$ has been chosen such
as to obtain a maximal stability of the sum rule in the region of interest.
As our range for $s_0$ we have chosen $s_0=4.2-5.8\;\gev^2$. The lower value
already lies close to the region of the $K_0^*(1950)$ resonance and around
the higher value the third scalar resonance would be expected from Regge
phenomenology. Thus the chosen range should be rather conservative and it is
gratifying that the most stable sum rule is reached for an $s_0$ within this
range. The dotted lines in figure \ref{fig1} show the corresponding variation
of $m_s$ with $s_0=4.2\;\gev^2$ (upper line) and $s_0=5.8\;\gev^2$ (lower line).
Again, the error on $m_s$ from the variation of $s_0$ is listed in table
\ref{tab1}. Because there is no stability for $s_0=4.2\;\gev^2$, as the
relevant value we have taken $m_s$ in the region around $1.6\;\gev$, where
stability occurs for the central parameters. For $\sqrt{u}\geq 2\;\gev$, the
continuum is larger than 50\% of the lhs of eq.~\eqn{sr}, and there the sum
rule becomes unreliable.

Except for the quark condensate, which will be discussed in the next section,
the values of the condensate parameters have been taken according to ref.
\cite{jm:95}. However, as already stressed above, their relevance for the
$m_s$ determination is negligible and thus also the corresponding uncertainty.
Instanton contributions to the scalar and pseudoscalar two-point functions
have been considered in refs. \cite{shu:83,dk:90,gn:93,deks:97,ess:98}. In the
framework of the instanton-liquid-model \cite{ss:98}, in ref.~\cite{km:01} it
was shown that the prediction for $m_s$ from scalar Borel sum rules is only
lowered by $2\;\mev$. In view of the uncertainties from other sources, we have
therefore neglected instanton contributions.

Adding the errors of table \ref{tab1} in quadrature, we arrive at our final
result for the strange quark mass:
\begin{equation}
\label{ms}
m_s(2\,\gev) \;=\; 99.4^{+16.1}_{-13.5} \;\mev \;=\; 99 \pm 16 \;\mev \,.
\end{equation}
To be more conservative, we have taken the larger of the errors as our final
uncertainty for the strange quark mass.

In ref.~\cite{jm:95,jam:98} the strange mass was also calculated from
the first moment sum rule which arises by differentiating eq.~\eqn{sr} with
respect to $u$. Performing this exercise here, we find that the resulting sum
rule is less stable and the region of maximal stability is lowered to about
$1\;\gev$, where perturbative as well as power corrections are more important.
Nevertheless, for our central value of $s_0$, $m_s$ only decreases by less than
$3\;\mev$, and if a lower $s_0$ is chosen to get a more stable sum rule, the
resulting value for $m_s$ is in complete agreement with eq.~\eqn{ms}, providing
additional support to our result.

\begin{figure}[htb]
\centerline{
\rotate[r]{
\epsfysize=15cm\epsffile{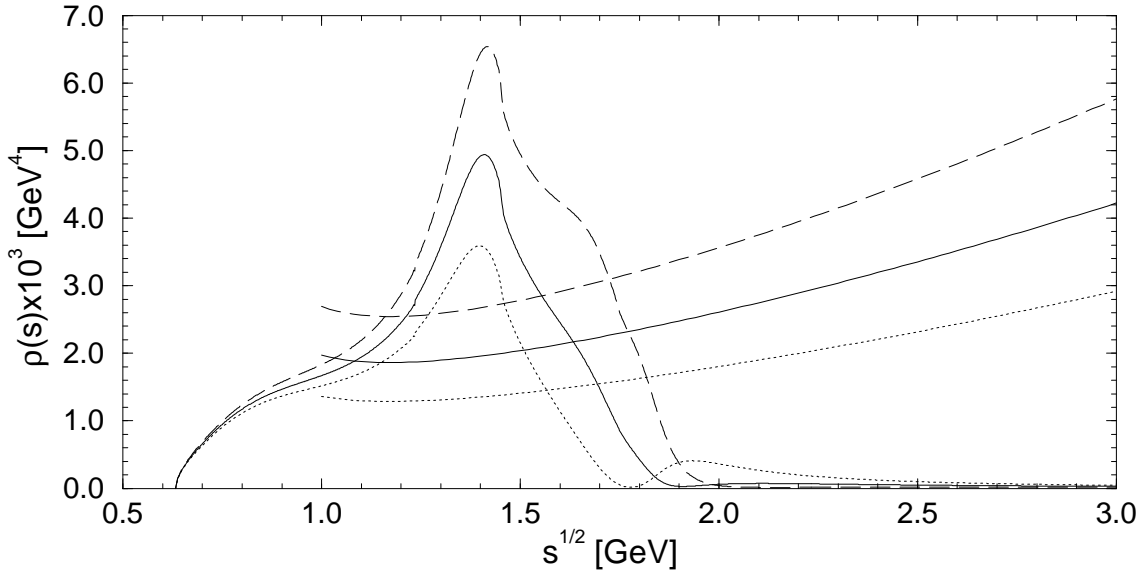} } }
\vspace{-4mm}
\caption[]{The theoretical as well as phenomenological spectral functions used
in our $m_s$ determination. Solid lines: central spectral functions; long-dashed
lines: $\rho_{ph}(s)$ for (6.10K4) with $F_{K\pi}(\Delta_{K\pi})=1.23$ and
$\rho_{th}(s)$ for $m_s=115\;\mev$; dotted lines: $\rho_{ph}(s)$ for (6.10K3)
with $F_{K\pi}(\Delta_{K\pi})=1.21$ and $\rho_{th}(s)$ for $m_s=83\;\mev$.
\label{fig2}}
\end{figure}

To conclude this section, in figure~\ref{fig2}, we display a comparison of the
theoretical as well as phenomenological spectral functions used in our $m_s$
determination. The solid lines correspond to central spectral functions. The
long-dashed lines show $\rho_{ph}(s)$ and $\rho_{th}(s)$ corresponding to the
largest value of $m_s$, and the dotted lines to the smallest. In the
$K_0^*(1430)$ resonance region, the hadronic spectral functions differ by
almost a factor of two. Thus, if it would become possible to experimentally
measure the scalar spectral function or $F_{K\pi}(s)$ in this region with
smaller uncertainties, the strange mass determination from scalar sum rules
could still be improved.

\newsection{Conclusions}

Let us now come to a comparison of our result \eqn{ms} for the strange quark
mass with other recent determinations of this quantity. A related approach to
the one followed here, also using the scalar sum rule, has been applied in
ref.~\cite{cfnp:97}, where $m_s(2\,\gev)=107\pm 13\;\mev$ was obtained. In this
work, however, the hadronic spectral function $\rho_{ph}(s)$ was estimated from
the single-channel Omn\`es form factor $F_{K\pi}^{\mbox{\tiny Omn\`es}}(s)$. In
view of our discussion about the dependence of the scalar $K\pi$ form factor
on the parametrisation of the corresponding S-wave $I=1/2$ phase shift in the
elastic, single channel case \cite{jop:01a}, the error in \cite{cfnp:97}
appears underestimated, although the central values are in good agreement.

The older scalar sum rule analyses of refs. \cite{jm:95,cdps:95,cps:96}, on
the other hand, have parametrised the phenomenological spectral function with
a Breit-Wigner Ansatz which was normalised to the scalar form factor at the
$K\pi$ production threshold. As was discussed in detail in refs. \cite{bgm:98,
cfnp:97}, this parametrisation overestimates the scalar spectral function,
because in the scalar channel the resonance contribution interferes
destructively with the large non-resonant background. Therefore, the resulting
strange mass values turned out larger than our central result presented here,
and should be discarded in the future. Nevertheless, within the uncertainties
at the time, including ${\cal O}(\alpha_s^3)$ corrections the result $m_s(2\,
\gev)=130\;\mev$ \cite{jm:95} still was compatible with our present finding of
eq.~\eqn{ms}.

Very recently, the determination of the strange mass from pseudoscalar finite
energy sum rules was reanalysed in ref.~\cite{km:01}. In this case, instanton
contributions play some role and have to be included. The resulting value
$m_s(2\,\gev)=100\pm 12\;\mev$ then is in perfect agreement to our finding of
eq.~\eqn{ms}. The status of the extraction of $m_s$ from the hadronic $e^+e^-$
cross section is less clear. Whereas ref.~\cite{nar:99} finds a value of
$m_s(2\,\gev)=129\pm 24\;\mev$, in \cite{mal:98} it is pointed out that large
isospin breaking corrections significantly lower the result for the strange
mass to about $m_s(2\,\gev)=95\;\mev$ and yield considerably larger
uncertainties of the order of $45\;\mev$. We therefore conclude that further
work in this channel is needed, before a definite conclusion can be reached.

The most recent determination of $m_s$ from the Cabibbo suppressed $\tau$-decay
width gave $m_s(2\,\gev)=116^{+20}_{-25}\;\mev$ \cite{cdghpp:01}, in agreement
with \eqn{ms} within the quoted error bars, although yielding a somewhat larger
central value. In addition to experimental uncertainties and a sizeable
sensitivity to the quark-mixing parameter $V_{us}$,\footnote{The reader should
note that a slightly lower central value for $m_s$ is obtained if the value
$|V_{us}|=0.2207$ \cite{lr:84,cknrt:01}, and not the unitarity-constraint
fit $|V_{us}|=0.2225$ \cite{pdg:00}, is used in the $\tau$ sum rule.} the
precision of the $\tau$-decay value is limited by the bad perturbative
behaviour of the $J=0$ contribution. Our determination of the scalar spectral
function could be used to disentangle the $J=0$ and $J=1$ components of the
$\tau$ data, allowing for a more accurate determination of $m_s$ from the
theoretically well behaved $J=1$ contribution.  In any case, whereas the
dominant uncertainty for $m_s$ from scalar sum rules arises from the
phenomenological part, in the $\tau$ decays it is due to the perturbative
expansion, and in this sense both determinations can be considered as
complementary.

Two recent reviews of determinations of the strange quark mass from lattice QCD
have been presented in refs.~\cite{lub:00,gm:01}, with the conclusions $m_s(2\,
\gev)= 110\pm 25\;\mev$ and $m_s(2\,\gev)= 120\pm 25\;\mev$ respectively. The
error in these results is dominated by the uncertainty resulting from dynamical
fermions, whereas the calculations of $m_s$ in the quenched theory, based for
example on the Kaon mass, are already very precise. Generally, in unquenched
calculations the strange mass is found below $100\;\mev$. Nevertheless, the
agreement between lattice QCD and QCD sum rule determinations of $m_s$ is
already very satisfactory.

Chiral perturbation theory provides rather precise information on ratios of
the light quark masses. Two particular ratios are \cite{leu:96}:
\begin{equation}
\label{mqrat}
R \;\equiv\; \frac{m_s}{\hat m} \;=\; 24.4\pm 1.5
\qquad \mbox{and} \qquad
Q^2 \;\equiv\; \frac{(m_s^2-\hat m^2)}{(m_d^2-m_u^2)} \;=\; (22.7\pm 0.8)^2 \,.
\end{equation}
From these ratios, one further deduces $m_u/m_d=0.551\pm 0.049$ and
$m_s/m_d=18.9\pm 1.3$, where the uncertainties have been estimated by assuming
Gaussian distributions for the input quantities. Our central values are in
agreement with the results quoted in \cite{leu:96}, although we find somewhat
larger errors. The ratio $m_u/m_d$ has also been calculated in \cite{abt:01}
with the result $m_u/m_d=0.46\pm 0.09$. Within the uncertainties, this ratio
is compatible with the previous one. Using the former ratios, together with
our result \eqn{ms} for $m_s$, we obtain for $m_u$ and $m_d$:
\begin{equation}
\label{mumd}
m_u(2\,\gev) \;=\; 2.9 \pm 0.6 \;\mev
\qquad \mbox{and} \qquad
m_d(2\,\gev) \;=\; 5.2 \pm 0.9 \;\mev \,.
\end{equation}
The resulting value for the sum of up and down quark masses, $(m_u+m_d)(2\,
\gev)=8.1\pm 1.4\;\mev$ is compatible with the finding $(m_u+m_d)(2\,\gev)=
9.6\pm 1.9\;\mev$ \cite{bpr:95,pra:98}, and in good agreement with the result
$(m_u+m_d)(2\,\gev)=7.8\pm 1.1\;\mev$ \cite{km:01}, both obtained with finite
energy sum rules for the pseudoscalar channel.

The knowledge of the light quark masses also allows for a determination of the
light quark condensate from the Gell-Mann-Oakes-Renner relation \cite{gmor:68}:
\begin{equation}
\label{gmor}
(m_u+m_d)\langle\Omega|\bar qq|\Omega\rangle \;=\;
-\,f_\pi^2 M_\pi^2\,( 1 - \delta_\pi ) \,.
\end{equation}
The term $\delta_\pi$ summarises higher order corrections in the chiral
expansion and also contains the renormalisation dependence mentioned in the
introduction \cite{jam:02}. Using a generous range $\delta_\pi=0.05\pm 0.05$
for this quantity \cite{jam:02}, together with the quark masses of eq.
\eqn{mumd} as well as $f_\pi=92.4\;\mev$ and $M_\pi=138\;\mev$, we arrive at:
\begin{equation}
\label{qq}
\langle\Omega|\bar qq|\Omega\rangle(2\,\gev) \;=\; -\,(267\pm 17\;\mev)^3 \,,
\end{equation}
which can be considered as an update of previous determinations of the light
quark condensate $\langle\bar qq\rangle$. Since it is still more common to
quote the quark condensate at a scale of $1\;\gev$ we also provide the
corresponding value: $\langle\bar qq\rangle(1\,\gev)=-\,(242\pm 16\;\mev)^3$.
This value can be compared with direct determinations of the quark condensate
from QCD sum rules \cite{dn:98}.

To conclude, in this work we have determined the masses of the light up, down
and strange quarks. To this end, first the strange mass $m_s$ was evaluated
in the framework of QCD sum rules for the scalar correlator with the result
\eqn{ms}. Our work improves previous analyses of this system by calculating
the phenomenological spectral function which enters the sum rule through a
dispersive coupled-channel analysis of the contributing hadronic states, making
use of our recent work \cite{jop:01a} on strangeness-changing scalar form
factors. The masses of the up and down quarks $m_u$ and $m_d$ were then
calculated employing ratios of quark masses known from $\chi$PT, together with
our result \eqn{ms} for $m_s$. Our final values for $m_u$ and $m_d$ have been
presented in eq.~\eqn{mumd}.

\bigskip
\subsection*{Acknowledgements}
This work has been supported in part by the German--Spanish Cooperation
Agreement HA97-0061, by the European Union TMR Network EURODAPHNE
(ERBFMX-CT98-0169), and by DGESIC (Spain) under the grant no. PB97-1261
and DGICYT contract no. BFM2000-1326. M.J. would like to thank the Deutsche 
Forschungsgemeinschaft for support.

\newpage

\end{document}